\documentclass[twocolumn]{aastex62}
\usepackage{color,xcolor}

\graphicspath{{./}{figures/}}

\received{August 14, 2019}
\revised{October 23, 2019}
\accepted{for publication in ApJ: October 24, 2019}

\shorttitle{A spectacular jellyfish galaxy in A1758N}
\shortauthors{Kalita \& Ebeling}

\begin{document}

\title{Jellyfish: Resolving the kinematics of extreme ram-pressure stripping at $z\sim0.3$\footnote{Data presented herein were obtained at the W.M.\ Keck Observatory, which is operated as a scientific partnership among the California Institute of Technology, the University of California, and the National Aeronautics and Space Administration. The observatory was made possible by the generous financial support of the W.M.\ Keck Foundation.}\thanks{Based on observations made with the NASA/ESA Hubble Space Telescope, obtained at the Space Telescope Science Institute, which is operated by the Association of Universities for Research in Astronomy, Inc., under NASA contract NAS 5-26555. These observations are associated with programs GO-14096.}}

\correspondingauthor{Boris Kalita}
\email{boris.kalita@cea.fr}

\author{Boris S.\ Kalita}
\affil{Institute for Astronomy, University of Hawaii, 2680 Woodlawn Dr,\\ Honolulu, HI 96822, USA}
\affil{CEA, IRFU, DAp, AIM, Universit\'e Paris-Saclay, Universit\'e Paris Diderot,\\ Sorbonne Paris Cit\'e, CNRS, F-91191 Gif-sur-Yvette, France}
\author{Harald Ebeling}
\affil{Institute for Astronomy, University of Hawaii, 2680 Woodlawn Dr,\\ Honolulu, HI 96822, USA}

\begin{abstract}
We present and discuss results from the first spatially resolved kinematic study of ram-pressure stripping of a massive late-type galaxy at intermediate redshifts. Our target, the spectacular ``jellyfish" galaxy A1758N\_JFG1, was previously identified as a fast-moving member of the equal-mass merger A1758N ($z=0.28$) with a star-formation rate of 48 M$_\sun$ yr$^{-1}$, far above the galaxy main sequence. IFU data obtained by us unambiguously confirm ram-pressure stripping as the physical mechanism driving the optical morphology and high star-formation rate of this system by revealing extended [\ion{O}{2}]$\lambda$3727\AA\ emission up to 40 kpc (in projection) downstream, as well as an ordered radial-velocity field generated by (a) conservation of angular momentum of the interstellar gas stripped from the edge of the galactic disk and (b) drag forces exerted by the intra-cluster medium on the ``tentacles" of stripped material. We find no evidence of significant nuclear activity in A1758N\_JFG1, although an AGN might, at this early stage of the stripping process, be obscured by high column densities of gas and dust near the galactic core. Finally, our exploration of possible trajectories of A1758N\_JFG1 found solutions consistent with the notions (a) that the A1758N merger proceeds along an axis that is substantially inclined with respect to the plane of the sky and (b) that A1758N\_JFG1 participated in the merger, rather than having been accreted independently from the field.
\end{abstract}


\section{Introduction} \label{sec:intro}

Galaxy evolution, specifically the transformation from late- to early-type galaxies, has long been known to depend sensitively on environment. Since low relative velocities are conducive to collisions, galaxies inhabiting low-density environments evolve primarily through mergers; by contrast, ram-pressure stripping \citep{gunn72} and galaxy harassment \citep{moore96} are key drivers of evolution in groups and clusters of galaxies where environmental effects dominate  \citep[e.g.,][]{oemler74,dressler80,whitmore93,dressler97}.
Ram-pressure stripping (RPS) in particular has emerged as the perhaps most efficient evolutionary process through which galaxies can lose their gas content, a critically important element of the transformation of gas-rich spirals to gas-devoid ellipticals. 

For galaxies traversing the dense, gaseous environment of clusters, the physics of RPS go well beyond the interplay between an external pressure gradient and a restoring gravitational force. In addition to the effects of turbulence and viscous stripping \citep[e.g.,][]{roediger08}, the large-scale kinematics of the interstellar medium (ISM) need to be considered if we want to understand the gradual degradation of spiral-arm features \citep{bekki02} or the sequence of stripping, currently understood to start with the loosely bound peripheral ISM, mainly HI \citep{haines84, cayatte90, vollmer01, kenney04, chung09, kenney14, kenney15, jaffe15, jaffe16}. Whether the molecular gas detected in the wake of galaxies affected by RPS was similarly removed from the disk is much more uncertain; a competing hypothesis posits that it formed \textit{in situ} through collapse of atomic gas clouds \citep{vollmer08,fumagalli09,boselli14,jachym14, moretti18}. 

Observations at UV and optical wavelengths of ionized gas that reveal the locations and rate of star formation represent a powerful tool in this context, as they highlight regions in which the collapse of gas clouds during stripping leads to the formation of young stars  \citep{cortese07,yoshida08,hester10,owers12,kenney14, fumagalli14,bellhouse17,jaffe18,vulcani18,poggianti19}. It remains debated though, both observationally and in numerical simulations, whether RPS-triggered star formation adds significantly (or, in fact, at all) to an increase in the galaxy's overall star formation rate \citep[SFR;][]{kronberger08,kapferer09,tonnesen12, roediger14}, and whether any enhancement of the star-formation rate occurs primarily in the disk or in the wake of stripped galaxies.

\begin{figure}
    \centering
    \includegraphics[width=0.45\textwidth]{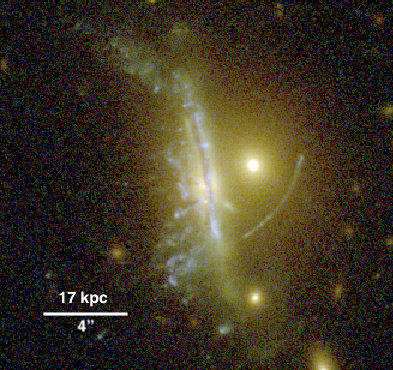}
    \caption{A1758N\_JFG1, a newly discovered spectacular example of ram-pressure stripping in a galaxy cluster at $z>0.2$ as seen with HST / ACS. Our target is viewed projected onto a cluster elliptical and an associated giant gravitational arc, both visible just west of this textbook ``jellyfish" galaxy.}
    \label{fig:a1758n-jfg1}
\end{figure}

By revealing the kinematics of the ISM, in both disk and wake of galaxies undergoing RPS, integral-field unit (IFU) observations of RPS cases in the nearby Universe have provided key insights into the complex physical processes governing ram-pressure stripping \citep[e.g][]{merluzzi13, fossati16, bellhouse17, gullieuszik17, pog17, vulcani18, fossati19}. Extending IFU studies of RPS to higher redshifts, we here attempt a detailed IFU-based characterization of A1758N\_JFG1  (Fig.~\ref{fig:a1758n-jfg1}), a spectacular ``jellyfish" galaxy discovered in our recent study of A1758N, an exceptional cluster merger at $z=0.28$ \citep[][hereafter EK19]{ebeling19} , as a first step toward a more comprehensive understanding of the impact of cluster environment and relaxation state on RPS events at ever increasing redshifts. 

Our paper is structured as follows: after introducing A1758N\_JFG1 in Sec.~\ref{sec:target}, we summarize the observations upon which our work is based in Sec.~\ref{sec:obs}. Data analysis procedures are discussed in Sec.~\ref{sec:analysis}, and results presented in Sec.~\ref{sec:results}. We then describe our simple three-body model of the gravitational interactions between A1758N and A1758N\_JFG1 in Sec.~\ref{sec:sim}, before interpreting and discussing our findings in Sec.~\ref{sec:discussion}. A summary is provided in Sec.~\ref{sec:summary}. Throughout this paper we adopt the concordance $\Lambda$CDM cosmology, characterized by  $\Omega_{m}=0.3$, $\Omega_{\Lambda}=0.7$, and $H_{0}=70$ km s$^{-1}$ Mpc$^{-1}$. All images are oriented such that north is up and east is to the left.

\begin{figure*}
    \centering
    \includegraphics[width=0.9\textwidth]{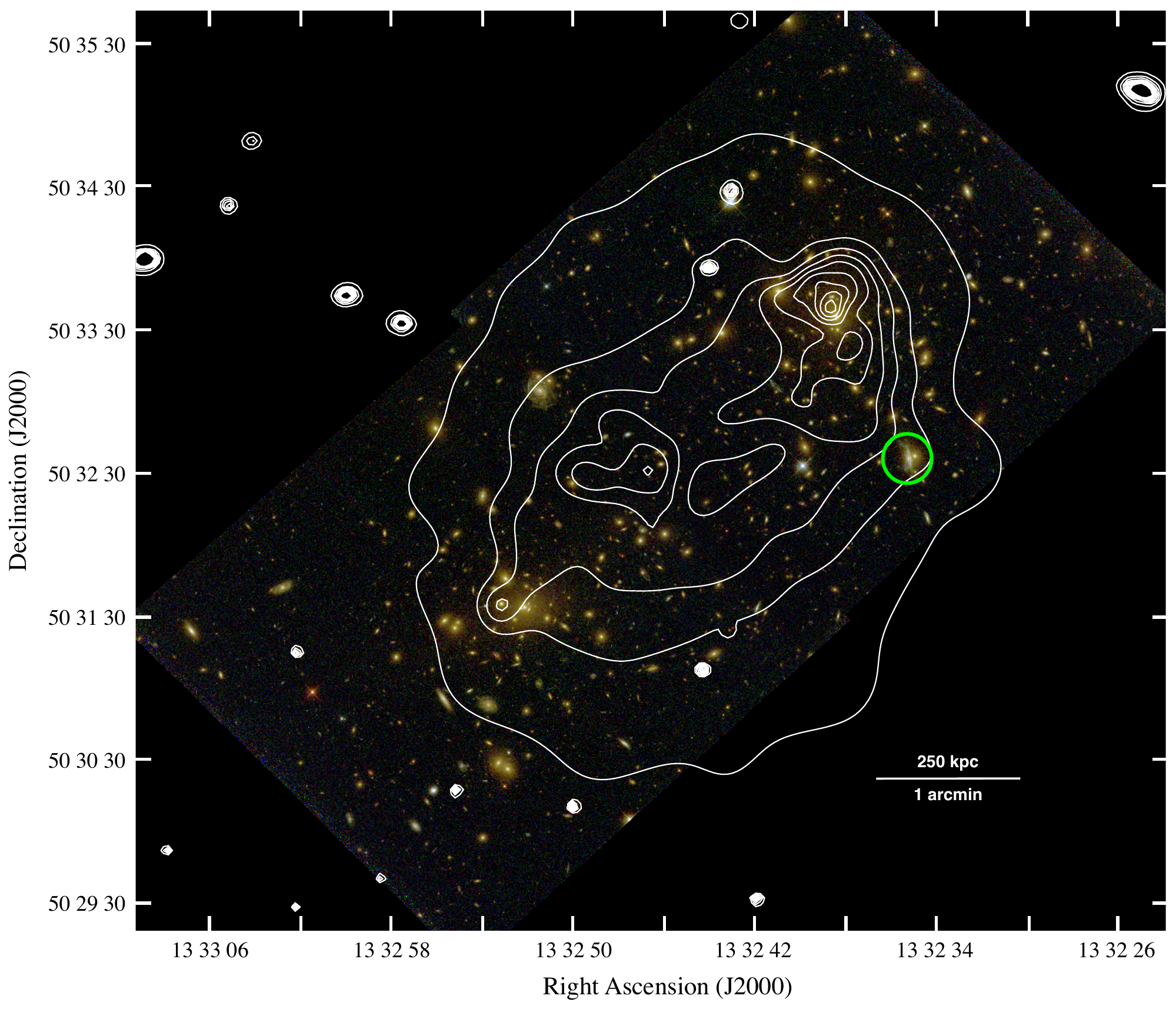}
    \caption{A1758N as seen with HST / ACS (false-colour composite from F435W, F606W, and F814W images collected for GO-12253). Overlaid in white are linearly spaced contours of the adaptively smoothed X-ray surface brightness in the 0.5--7 keV band as observed with Chandra / ACIS-I. A1758N\_JFG1 is marked by a green circle.}
    \label{fig:a1758-hst}
\end{figure*}

\section{A spectacular case of ram-pressure stripping}
\label{sec:target}
The target of this work, A1758N\_JFG1 (Fig.~\ref{fig:a1758n-jfg1}) was first noted by \citet[][their Fig.~3]{ragozzine12} in the context of their weak-lensing analysis of the A1758 double cluster \citep{abell58}. However, focusing entirely on strong-lensing features, these authors made no mention of the spectacular ``jellyfish" morphology \citep{smith10,ebeling14,mcpartland16} of the giant, edge-on spiral which they (most likely erroneously) take to be the lens responsible for the giant arc visible in Fig.~\ref{fig:a1758n-jfg1}\footnote{ \citet{ragozzine12} describe this source as  ``a good candidate [...] for being a strong arc and for being lensed by the edge-on spiral to the east".}. A1758N\_JFG1 was re-discovered by us in 2018 in the course of a visual search for RPS candidates in A1758N, described in detail in EK19.

\section{Observational Data}
\label{sec:obs}

Our analysis is based on data previously used by EK19 for their study of RPS events in A1758N, as well as on IFU observations performed specifically for this work. We here provide a brief overview of the former before describing the latter in more detail. 

\subsection{Space-based optical and X-ray observations}
\label{sec:hst-obs}

Observations of A1758N were performed with the Hubble Space Telescope (HST) for two unrelated projects (GO-12253, PI: Clowe, and GO-14096, PI: Coe) in 2011 and 2016, respectively. We used the resulting high-level science products from the Advanced Camera for Surveys (ACS) publicly available from the MAST archive and refer to EK19 for more details about these observations. A1758 was also observed twice with the Chandra X-ray Observatory (Sequence Number 800152 and Sequence Number 801177, PI: David) in 2001 and 2012, respectively. We only used the data from the second, much deeper observation and again refer to EK19 for additional details. An overlay of the adaptively smoothed X-ray emission \citep{ebeling06} on the optical HST image is shown in Fig.~\ref{fig:a1758-hst}.

\subsection{Ground-based spectroscopy}
\label{sec:spec}
\subsubsection{Long slit}

We observed A1758N\_JFG1 (and other RPS candidates in A1758N, see EK19) with the DEIMOS multi-object spectrograph on the Keck-II 10m-telescope in poor conditions in July 2018. The chosen instrumental setup used a 1\arcsec\ slit width and combined the 600 l/mm grating (set to a central wavelength of 6300\AA) with the GG455 blocking filter to suppress second-order contributions at $\lambda>9000$\AA.

\subsubsection{Integral Field Unit (IFU)}

About a year later, on April 29, 2019, we targeted A1758N\_JFG1 again, this time with the Keck Cosmic Web Imager \citep[KCWI;][]{morrissey18}. In order to cover the entire disk of the galaxy as well as the downstream region while maintaining adequate spatial resolution, we chose the Medium Slicer mode, which provides 0.7\arcsec\ spatial sampling and a 16.5\arcsec$\times$20\arcsec\ field of view, well matched to the 19\arcsec$\times$19\arcsec region shown in Fig.~\ref{fig:a1758n-jfg1}. Since KCWI's spectral coverage is currently limited to 3500 to 5600\AA, we opted for the BM grating and a central wavelength of 5000\AA, thereby covering the range from 4550 to 5450\AA\ (approximately 3600--4300\AA\ in the galaxy restframe). This spectral window includes the redshifted location of the [\ion{O}{2}]3727\AA\ emission line, an important star-formation diagnostic. We integrated for a total of 260 minutes in variable, but consistently poor seeing of 1.2 to 2\arcsec. All data were reduced with the python-based data reduction pipeline \textsc{KDERP}.

\section{Data Analysis}
\label{sec:analysis}

\subsection{Photometry}
\label{sec:phot-data}
\begin{figure}
    \centering
    \hspace*{-5mm}\includegraphics[width=0.5\textwidth]{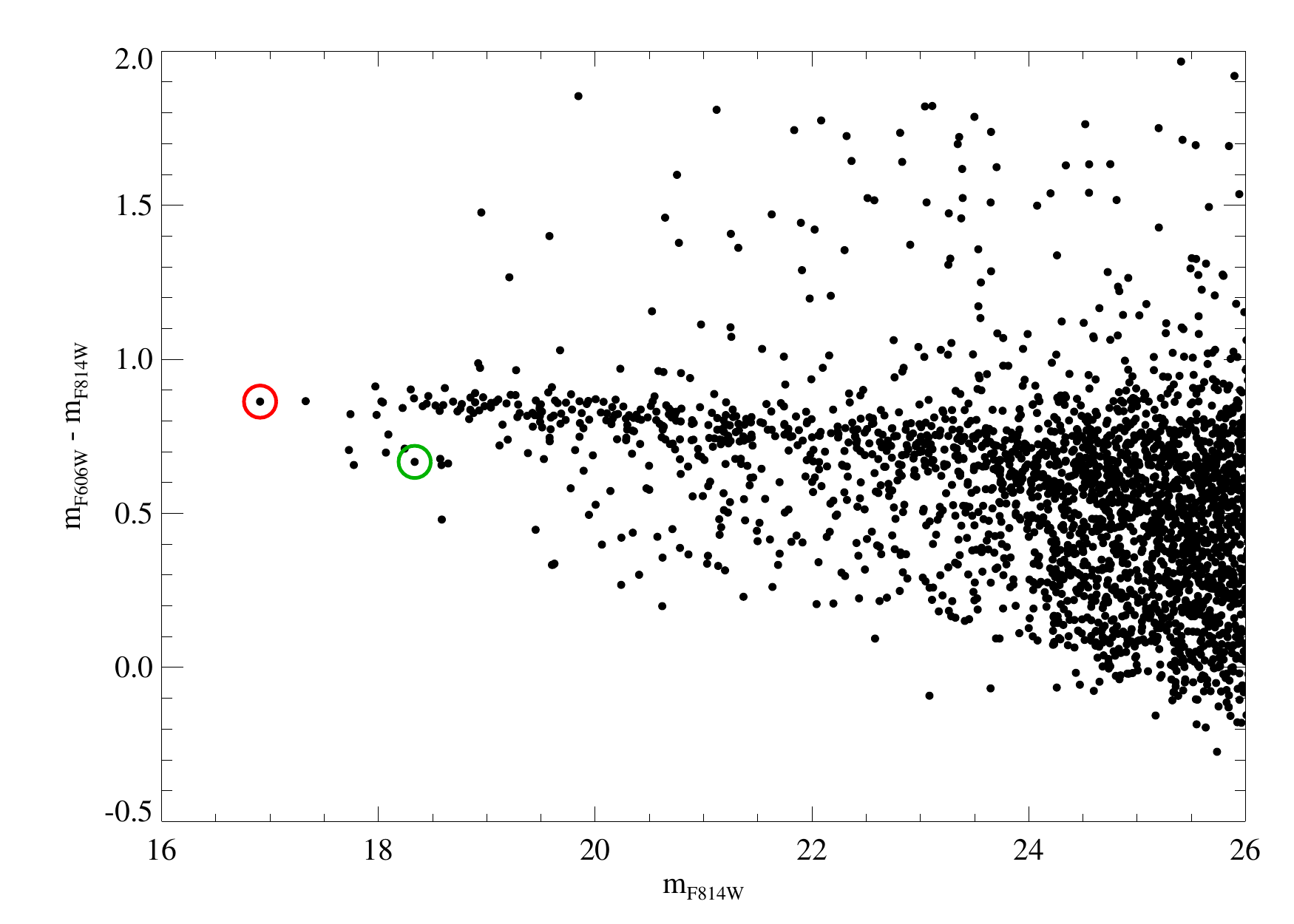}
    \caption{Color-magnitude diagram of all galaxies within the field of Fig.~\ref{fig:a1758-hst}. A1758N\_JFG1 and the BCG NNE of it (Fig.~\ref{fig:a1758-hst}) are marked by a green and red circle, respectively.}
    \label{fig:cmd}
\end{figure}

We initially obtained photometry in all HST passbands with SExtractor \citep{bertin96} in dual-image mode and using the F606W image as the detection band, while adopting the same settings and input parameters as employed for GO-14096.
Since the resulting photometry proved inadequate for the complex superposition of sources in the immediate vicinity of A1758N\_JFG1 (see Fig.~\ref{fig:a1758n-jfg1}), we obtained custom photometry  for this object by subtracting a GALFIT model of the nearby elliptical galaxy (visible to the West in Fig.~\ref{fig:a1758n-jfg1}) and then adjusting SExtractor's parameters to avoid fragmentation of the object of interest, the edge-on massive spiral. Differential photometry was also applied in order to subtract any remaining contribution from the much fainter elliptical visible near the bottom of Fig.~\ref{fig:a1758n-jfg1}. The brightness and color of A1758N\_JFG1 relative to other galaxies within the HST/ACS field of view is illustrated by Fig.~\ref{fig:cmd}.

\subsection{Long-slit spectroscopy}
\label{sec:spec-cal}

\begin{figure}
    \centering
    \hspace*{-5mm}\includegraphics[width=0.49\textwidth]{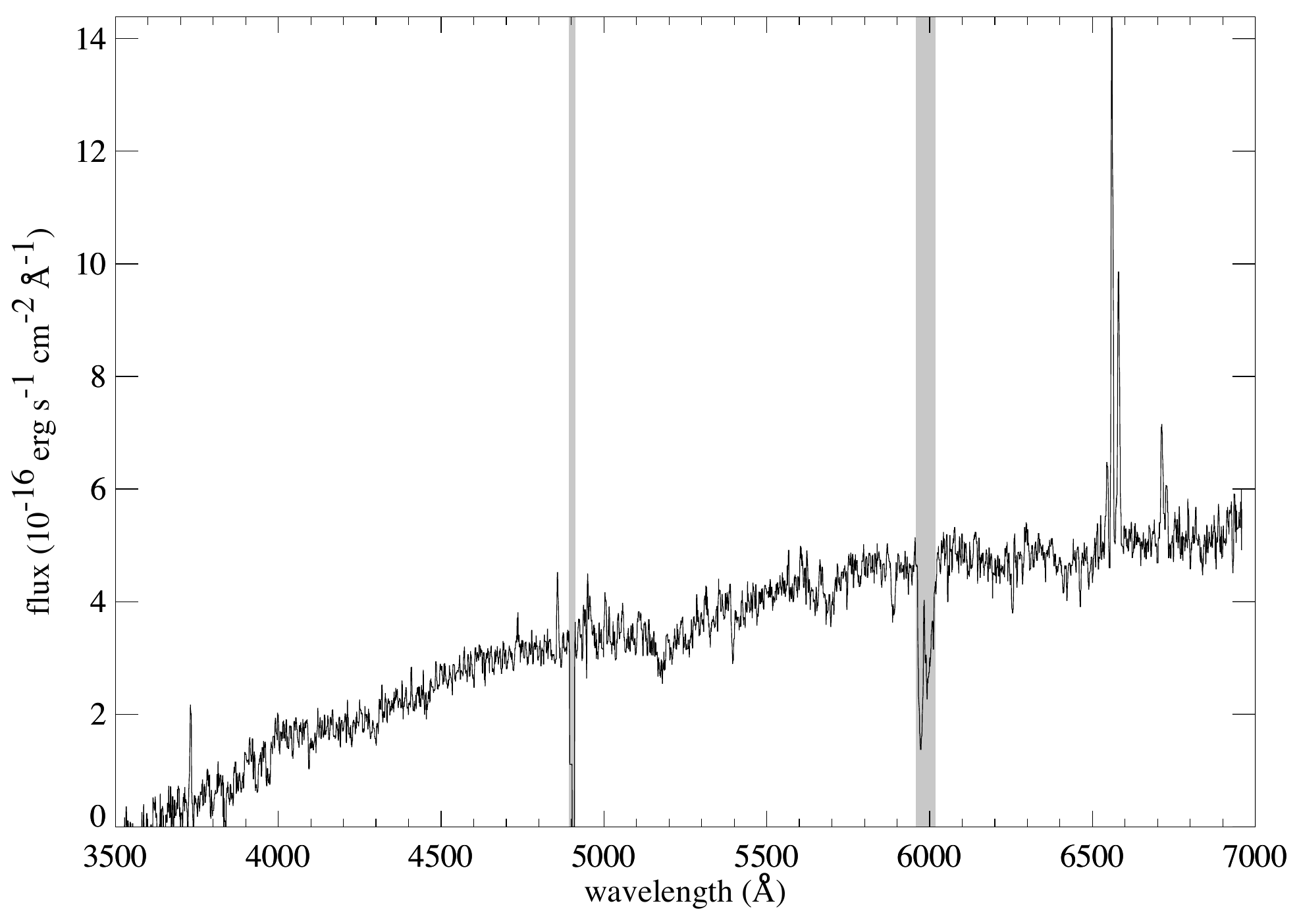}
    \caption{Rest-frame spectrum of A1758N\_JFG1 as obtained with Keck-II/DEIMOS. The DEIMOS chip gap and absorption at 7600\AA\ (about 6000\AA\ in the galaxy rest frame) from water in the atmosphere are greyed out.}
    \label{fig:1dspec}
\end{figure}

The DEIMOS spectrum was flux calibrated by tying the observed flux to the HST photometry within the same passband and was subsequently extinction corrected. The procedures employed are described in detail in EK19. Note that, as for all studies based on long-slit observations, any galaxy properties derived from the resulting spectrum, shown in Fig.~\ref{fig:1dspec}, implicitly assume that the signal recorded within the DEIMOS slit is representative of that of the entire galaxy. 

\label{sec:vel-an}
\begin{figure}
    \centering
    \includegraphics[width=0.45\textwidth]{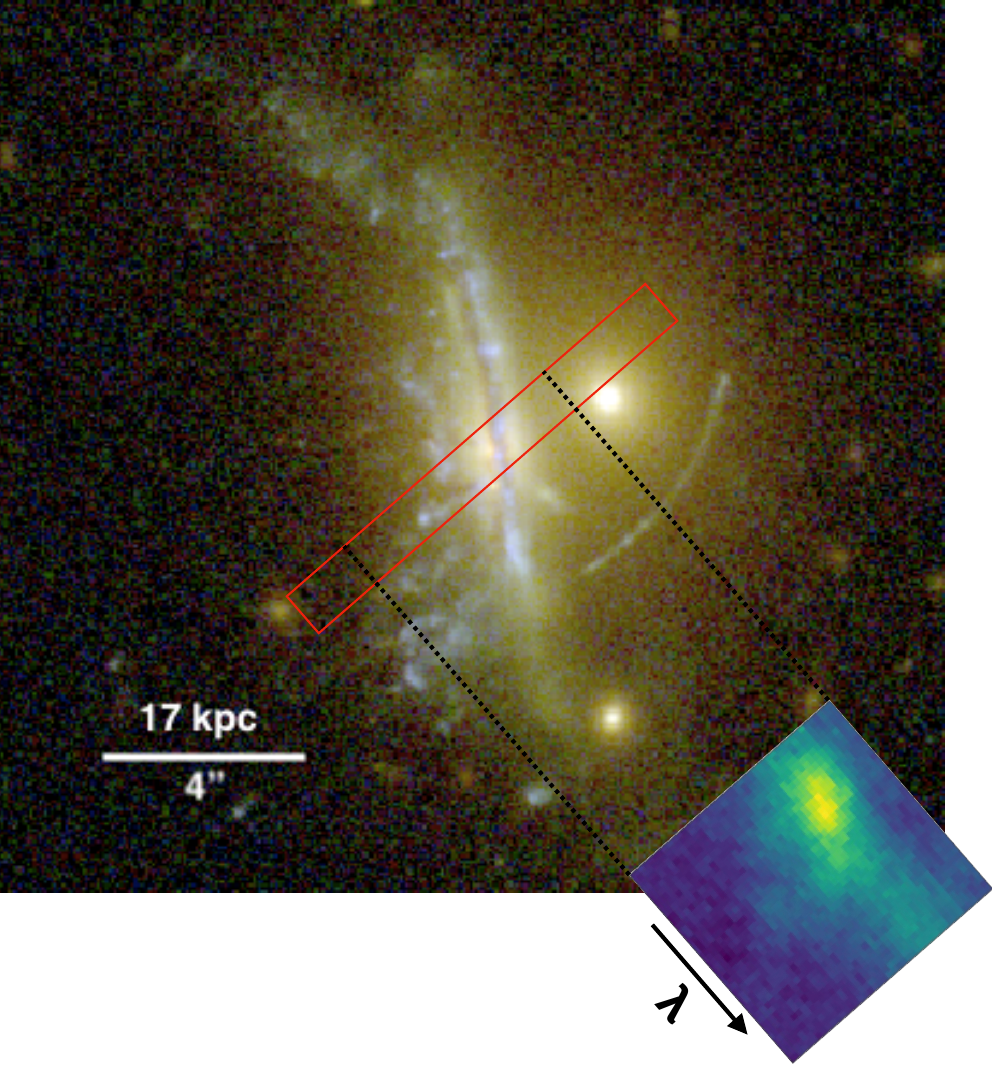}
    \caption{Stacked 2d spectrum of 20\AA-wide regions around each of  H$\alpha$, [\ion{N}{2}], H$\beta$ and [\ion{O}{3}]5007\AA, superimposed on the HST image of A1758N\_JFG1 to illustrate the spatial alignment of spectral and imaging features along the DEIMOS slit. Note the clear slant of the stacked emission lines.}
    \label{fig:a1758n-jfg1-spectrum}
\end{figure}

From the DEIMOS 2d spectrum of A1758N-JFG1 we measured the galaxy's rotational velocity as a function of galactocentric radius by stacking sections of the 2d spectra around the H$\alpha$, [\ion{N}{2}]6548\AA, [\ion{O}{3}]5007\AA, and H$\beta$ emission lines (Fig.~\ref{fig:a1758n-jfg1-spectrum}). Gaussian profiles were fitted to each of the spatial rows, and the centroids of these Gaussians converted to radial velocities using the Doppler-shift formula. Since the DEIMOS slit runs, by design, through the approximate center of the galaxy at an angle of 58$^\circ$ relative to the plane of the galactic disk, the galactocentric distance of any feature on the slit can be calculated via $d_{\rm g} = d_{\rm s} \cos 58^\circ$, where  $d_{\rm g}$ and $d_{\rm s}$ are the distance from the center of  A1758N\_JFG1 and the distance along the slit, measured from the center of the continuum emission.  

\subsection{IFU data} 

\begin{figure}
    \centering
    \hspace*{0mm}\includegraphics[width=0.45\textwidth]{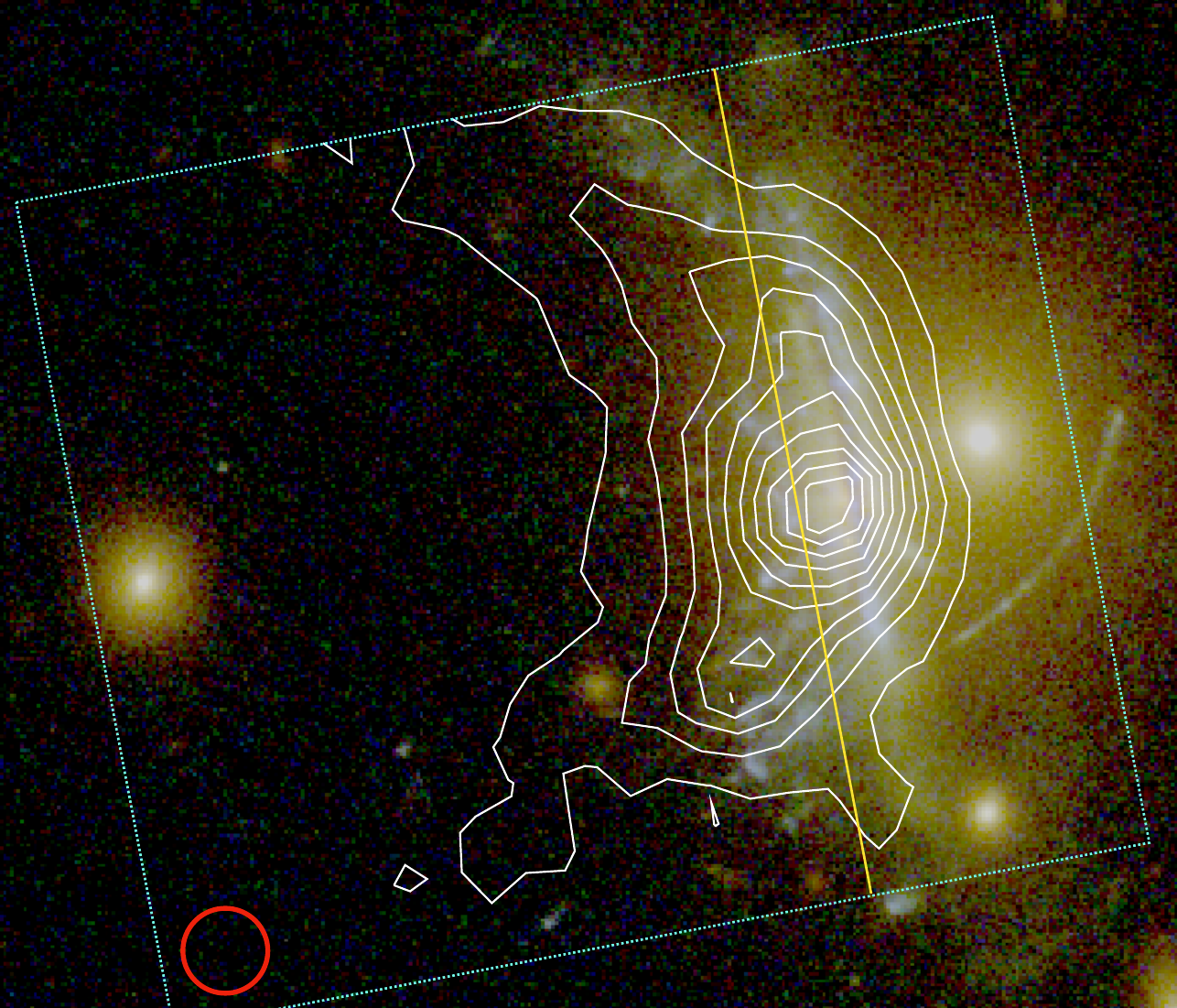}
    \caption{Contours (linearly spaced) of the [\ion{O}{2}]3727\AA\ emission-line flux from A1758N\_JFG1 overlaid on the HST image of our target show the stripping of gas from the outer edge of the disk. The dotted, cyan line delineates the KCWI field of view; the red circle has a diameter of 1.5\arcsec, representing the seeing disk during our Keck-II observation. The yellow line divides the field into a ``disk" and ``wake" region (see text for details).  }
    \label{fig:o2flux}
\end{figure}

As the first step toward analyzing the [\ion{O}{2}] emission from A1758N\_JFG1, we limited the spectral range of the fully reduced and calibrated KCWI data cube and the associated signal variance (the ``icubes" and ``vcubes" files, in KDERP parlance) to 80\AA\ centered on 4745\AA, the observed wavelength of [\ion{O}{2}] at the systemic redshift of our target, as determined from the DEIMOS long-slit spectrum shown in Fig.~\ref{fig:1dspec}. We also binned the cubes by a factor of three in both spatial dimensions to create 0.87\arcsec\ pixels, approximately matched to the 0.7\arcsec\ slit width. We then computed the net [\ion{O}{2}] flux at each pixel location as the difference of the signal within a 14\AA-wide region centered on [\ion{O}{2}] and the sum of the (continuum) flux recorded in 7\AA-wide regions at either end of our spectral window. The resulting image of the [\ion{O}{2}] emission-line flux is shown in Fig.~\ref{fig:o2flux}.

\begin{figure*}
    \centering
    \hspace*{-5mm}\includegraphics[width=0.8\textwidth]{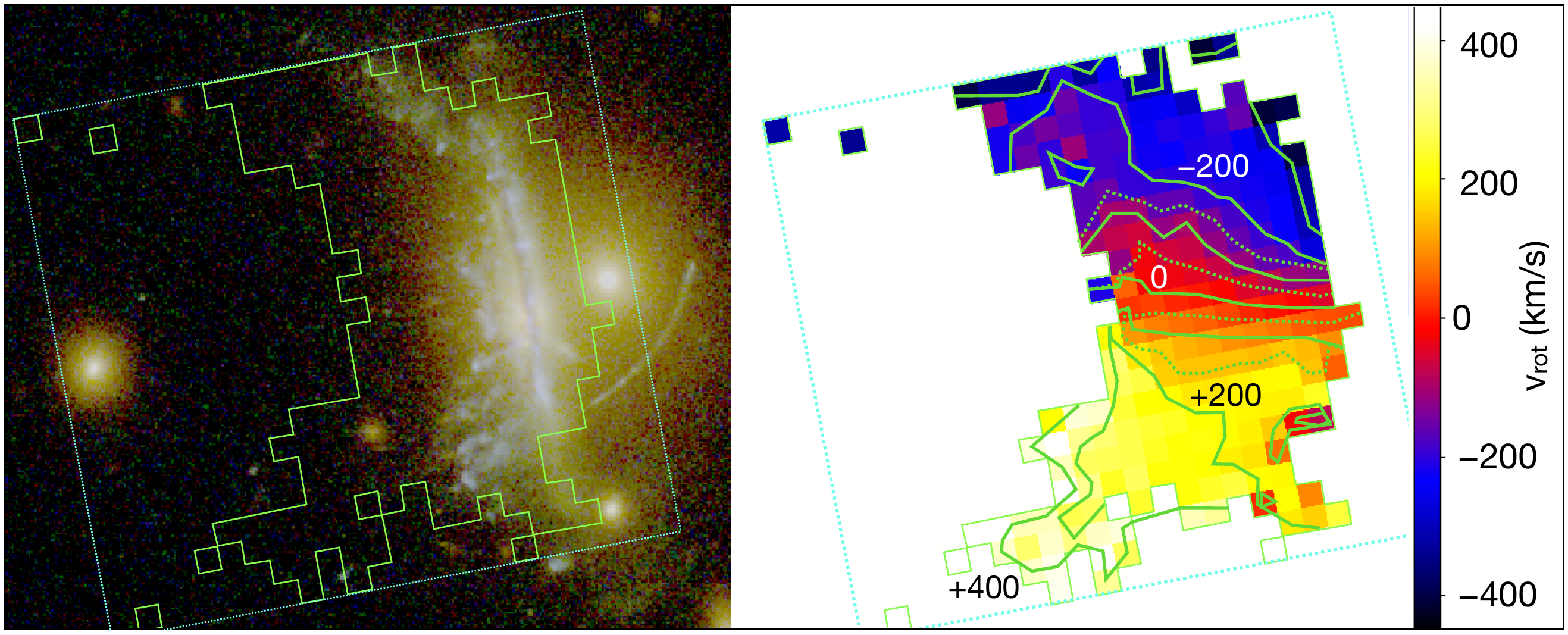}
    \caption{Right: map of the radial velocity of A1758N\_JFG1 as derived from the observed wavelength of the [\ion{O}{2}]3727\AA\ emission line. The dotted, cyan line delineates the KCWI field of view; within the shown outlines the line flux is significant at greater than 2$\sigma$ significance. Left: outlines of the $v_{\rm rot}$ distribution shown on the right, overlaid on the HST image of our target. Note that our measurement probes the velocity field well downstream of the disk, i.e., the jellyfish ``tentacles".  }
    \label{fig:vrotmap}
\end{figure*}

Finally, we measured radial velocities from the location of [\ion{O}{2}] in the spectral dimension of the data cube. In order to avoid systematic uncertainties introduced by fitting a Gaussian profile to the unresolved 
[\ion{O}{2}] doublet, we computed the line centroid as the flux-weighted average of the wavelengths within the same 14\AA-wide spectral window used before. The results, after conversion to peculiar velocities in the galaxy rest frame and considering only pixels within which [\ion{O}{2}] was detected with greater than 2$\sigma$ significance, are shown in Fig.~\ref{fig:vrotmap}.

\section{Results}
\label{sec:results}

\begin{table*}
\centering
\begin{tabular}{cccccc@{\hspace*{1mm}}c@{\hspace*{1mm}}c}
R.A.\ (J2000) Dec & $z$  & m$_{435}$ &  m$_{606}$ & m$_{814}$ & log(M$\ast$ M$_{\odot}^{-1}$) & SFR  & $\Delta v_{\rm rad}$ \\ 
&  & & & & & (M$_{\odot}$ yr$^{-1}$)  & (km s$^{-1}$)\\ \hline
13 32 35.18 +50 31 36.3 & 0.2733 & 19.70 & $\;18.58$ & $\;17.97$  & 10.87$\pm$0.02 &     47.9$\pm$11.9 &$\;\;-936$\\ \hline
\end{tabular}
\caption{Global properties of A1758N\_JFG1: equatorial coordinates, heliocentric redshift, apparent magnitudes in the F435W, F606W and F814W bands, respectively, stellar mass, star formation rate (SFR), and radial velocity relative to the NW BCG of A1758N. 
\label{tab:galz}}
\end{table*}

\subsection{Global properties}
\label{sec:globprop}

From the prominent emission lines shown in Fig.~\ref{fig:1dspec} EK19 measured a heliocentric redshift of $z=0.2733$ for A1758N\_JFG1, significantly less than the redshift of $z=0.2787$ of the nearby brightest cluster galaxy of the north-west component of A1758N (see Fig.~\ref{fig:a1758-hst}). As reported by EK19, this redshift difference corresponds to a relative radial velocity of about $-940$ km s$^{-1}$, i.e., in addition to its apparent westerly motion in the plane of the sky, A1758N\_JFG1 is moving rapidly toward us in the cluster rest frame.

The stellar mass of $(7.4\pm0.3)\times10^{10}$ M$_\odot$ and star-formation rate of $(48\pm12)$ M$_\odot$ yr$^{-1}$ measured for A1758N\_JFG1 by EK19 place our target far above the galaxy main sequence, about a factor of five above the star formation rates encountered typically in disk galaxies of this mass (EK19; their Fig.~8). Consistent with this observation, EK19 find A1758N\_JFG1 to fall within the starburst regime in the BPT diagram \citep{baldwin81}, albeit just outside the purely star-forming population (EK19; their Fig.~9). While not representative of the field galaxy population in general, the system's high stellar mass \citep[within the top few percent of disk galaxies at $z<0.5$;][]{torrey15} is accompanied by a commensurate physical size, absolute brightness, and resulting SFR under RPS (see Fig.~8 of EK19), rendering A1758N\_JFG1 an ideal target for in-depth study.

\begin{figure}
    \centering
    \hspace*{-1mm}\includegraphics[width=0.45\textwidth]{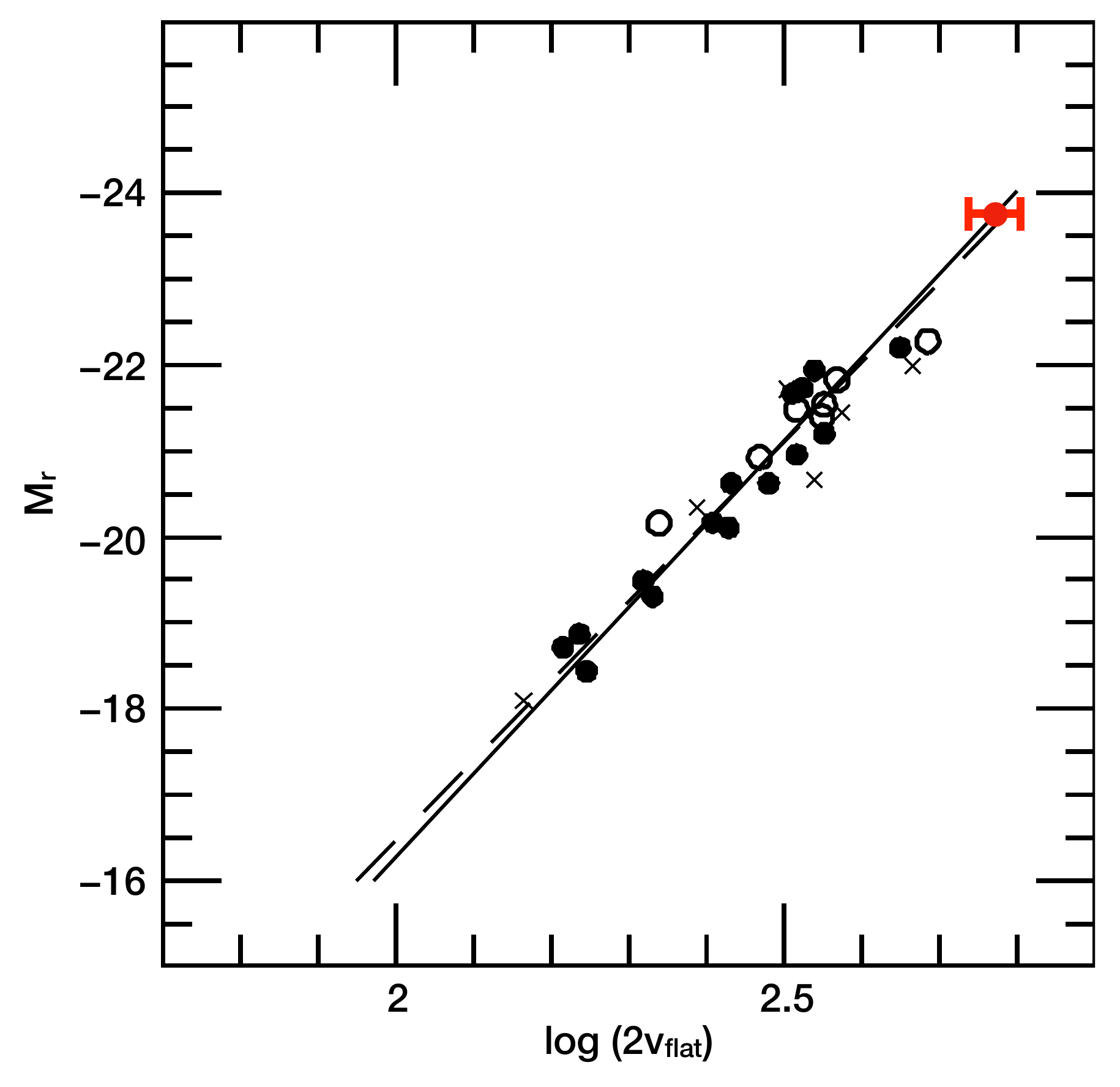}
    \caption{Locus of A1758N\_JFG1 (red symbol) on the R-band Tully-Fisher relation at $z=0$ \citep{verheijen01}. From the extinction-corrected absolute R-band magnitude of -23.7 (derived from the k-corrected F814W magnitude), the TFR predicts a rotation velocity of approximately 300 km s$^{-1}$ for our textbook ``jellyfish" galaxy.}
    \label{fig:tfr}
\end{figure}

Our analysis of the stacked 2D spectrum (Fig.~\ref{fig:a1758n-jfg1-spectrum}) from the same DEIMOS observation as used by EK19 yielded a maximal velocity for the ionized ISM of (143$\pm$16) km s$^{-1}$. For comparison, we also calculated the rotational velocity of A1758N\_JFG1 expected from the Tully-Fisher relation \citep[TFR;][]{tully77} as determined by \citet{verheijen01}. Since the F814W passband at z=0.28 is effectively R-band at $z=0$, we used our target's F814W magnitude, corrected for Galactic extinction (Sec.~\ref{sec:phot-data}), to find V$_{\rm flat}$ = (288$\pm$31) km s$^{-1}$ from the R-band TF relation\footnote{The effects of a possible redshift evolution of the TF relation \citep{port07} amount to at most 10\% at the modest redshift of our target and thus do not affect our conclusions.} (Fig.~\ref{fig:tfr}). The discrepancy between this expectation value and the cited, much lower measurement obtained from our DEIMOS observation is entirely anticipated, given that the spectrum recorded through our DEIMOS slit (see Fig.~\ref{fig:a1758n-jfg1-spectrum}) probes only the central region of the galaxy, far from the outer reaches of the disk where the rotational-velocity curve flattens. 

The global properties of A1758N\_JFG1 derived here (or previously by EK19) show our ``jellyfish" galaxy to be a massive, star-bursting, fast-rotating, late-type galaxy moving at high velocity through the A1758N cluster merger (Table~\ref{tab:galz}). Our target's direction of motion in the plane of the sky is, as for all galaxies, difficult to determine but can be approximately constrained from the orientation and slight N-S asymmetry of the ``tentacles" in Fig.~\ref{fig:a1758n-jfg1} to be due west or WSW.

\subsection{Nuclear activity and star-formation rate}
\label{sec:sfna}
Prompted by A1758N\_JFG1 falling into the composite region of EK19's BPT diagram, we attempted to constrain the fraction of the H$\alpha$ luminosity that can be attributed to LINER/AGN-powered photoionization. Although the absence of  diagnostic emission lines (other than [\ion{O}{2}]3727\AA) within the spectral window \citep[e.g.,][]{poggianti19} currently precludes a conclusive determination from our IFU data, an upper limit on any AGN contribution can be obtained from the lack of X-ray emission from A1758N\_JFG1, apparent in Fig.~\ref{fig:a1758-hst}. The 3$\sigma$ upper limit on the X-ray luminosity of $2\times 10^{42}$ erg s$^{-1}$ (2--10 keV) measured by us from the archival Chandra data suggests a negligible AGN contribution of less than $2\times 10^{41}$ erg s$^{-1}$ to the observed H$\alpha$ luminosity of 2$\times$10$^{42}$ erg s$^{-1}$ \citep{shi10}.

While the narrow spectral range covered by our KCWI data contains insufficient spectral diagnostics to shed further light on any nuclear activity, the IFU data provide a clear view of the spatial distribution of star formation. Adopting the [OII]3727\AA\ flux as an alternative tracer of star formation \citep[if less directly so than H$\alpha$;][]{ken98}, we find star formation to  occur not only along the entire galactic disk but also, at steadily declining intensity, well outside the disk, in the downstream region (Fig.~\ref{fig:o2flux}). Indeed the integrated SFR in the galaxy's wake (see dividing line in \ref{fig:o2flux}) rivals, and in fact exceeds, that observed in the disk region. We also note that the contours of the [OII]3727\AA\ flux shown in Fig.~\ref{fig:o2flux} exhibit the same asymmetry noted already in Sec.~\ref{sec:globprop} with reference to the orientation of the tentacles visible in Fig.~\ref{fig:a1758n-jfg1}.

\subsection{Gas phase velocities}

A conclusive, and much more complex, picture of the rotational velocity of A1758N\_JFG1 is painted by the map we derived from the redshift of the [\ion{O}{2}] emission line (Fig \ref{fig:vrotmap}). The key insights can be summarized as follows:

\begin{enumerate}
    \item All lines of constant radial velocity are significantly slanted relative to the galaxy's axis of rotation, in the sense that gas that is farther downstream (i.e., east) of the galactic disk lags behind the galaxy (which is moving toward us, see Sec.~\ref{sec:globprop}). 
    \item The lag is small near the center of A1758N\_JFG1, but increases both toward the outer edges of the disk and eastward, along the debris trail, where offsets of over 200 km s$^{-1}$ are observed.
    \item Within the galactic disk, the maximal rotational velocity observed once the mentioned lag is accounted for ($v_{\rm max}\sim\pm 300$ km s$^{-1}$) is consistent with the TFR expectation value.
\end{enumerate}

\section{Origin and trajectory within the A1758N merger}
\label{sec:sim}

From their analysis of the line-of-sight and projected directions of motions of eight RPS candidates in A1758N (one of them A1758N\_JFG1) EK19 concluded that these galaxies likely have a common origin, the most plausible scenario being infall along a wide filament roughly aligned with the cluster merger axis.

To test the validity of this hypothesis we attempted to recover the trajectory of A1758N\_JFG1 by creating a simple three-body model of the A1758N collision, constrained by five observational facts: the projected positions of the two BCGs and of A1758N\_JFG1, the line-of-sight velocity of our jellyfish galaxy, and its projected, approximately westerly direction of motion.

\begin{figure*}
    \centering
    \hspace*{0mm}\includegraphics[width=0.98\textwidth]{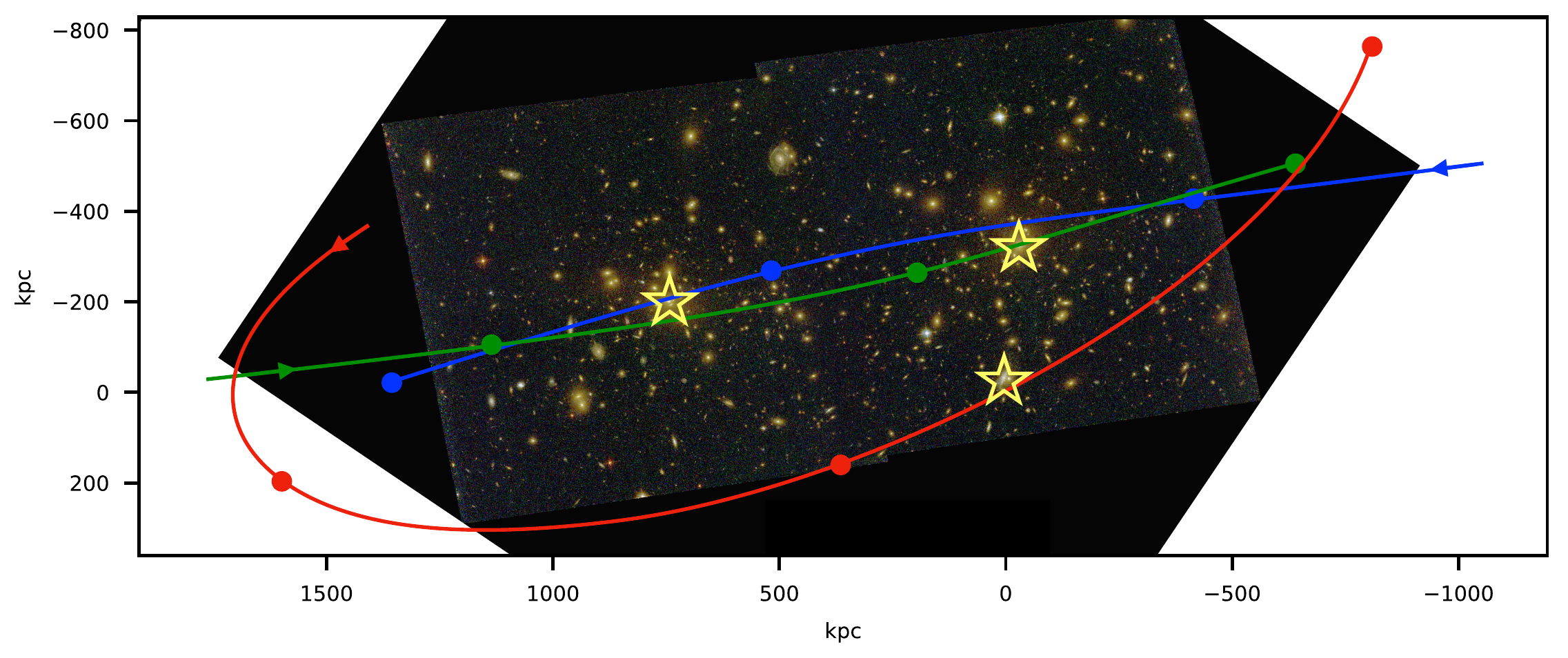}
    \caption{Example of simulated trajectories for the two subclusters and A1758N\_JFG1 that reproduce the observed locations (in projection) of the two BCGs and our jellyfish galaxy (marked by yellow asterisks) as well as the radial velocity and approximate projected direction of motion of the latter. We show the trajectories overlaid on the HST image of A1758N, i.e., as viewed along our line of sight; in three dimensions, the merger axis for which our observational constraints are met is inclined by 58$^\circ$ with respect to the plane of the sky. Solid circles on each trajectory mark time intervals of 667 Myr.}
    \label{fig:galpy}
\end{figure*}

Since realistic modelling of the complex physics of a cluster merger was well beyond the scope of our conceptual study of plausible orbits of A1758N\_JFG1, we simulated the relevant trajectories with \textsc{GALPY}, a python package designed for galactic-dynamics calculations \citep{bovy14, mack18}. The two subclusters in A1758N were assigned spherical NFW mass profiles \citep{nfw96} of 1.1$\times 10^{15}$ M$_\odot$ each \citep[based on findings of][]{ragozzine12}, whereas A1758N\_JFG1 was considered a point mass; in addition, all three bodies were assumed to share the same orbital plane. We adopted the initial conditions of \citet{machado15} and \citet{mo17}, i.e., we started our simulations from a subcluster separation of 3 Mpc and assumed an impact parameter of 150 kpc and a relative velocity of 1750 km s$^{-1}$ between the two clusters. The parameter space probed by the initial conditions for A1758N\_JFG1 combined distances ranging from 600 kpc to 2 Mpc from each of the two subclusters with an initial velocity vector ranging from 600 to 2400 km s$^{-1}$ and adopting all possible angles within the merger plane. Dynamical friction as well as hydrodynamical interactions were ignored, a decision that was as much choice as necessity, given our inability to incorporate such physics into GALPY. Neither simplification is likely to have a noticeable impact on our findings though, since the merger is believed to be observed after the first core passage \citep{machado15,mo17}, i.e., before dynamical friction has a noticeable effect, and that A1758N\_JFG1 appears to be in an early phase of the stripping process. For this last reason, and because the stripping timescale is shorter than the dynamical timescale of galaxies within clusters \citep{quilis00, steinhauser16}, we also required A1758N\_JFG1 not to have previously passed either of the two subcluster cores within a distance of 500 kpc, which is a conservative estimate of the pericenter of orbits of stripped galaxies within clusters \citep{quilis00}. We note that this is the only constraint used that at least implicitly invokes the density of the intra-cluster medium; as stated before, a realistic simulation of the gas densities encountered by A1758N\_JFG1 on its trajectory through this active merger is well beyond the scope of our simple three-body simulation.

The set of observational constraints described above greatly limit the possible trajectories of A1758N\_JFG1. Although our simple simulations are far from exhaustive, the results favor a scenario in which the A1758N merger proceeds along an axis that is closer to our line of sight than to the plane of the sky, and in which A1758N\_JFG1's trajectory is not dissimilar to that of the subcluster observed today as the NW component of A1758N (see Fig.~\ref{fig:galpy}).

\section{Discussion}
\label{sec:discussion}
The discovery of A1758N\_JFG1, a spectacular example of ram-pressure stripping in a massive cluster merger, provides a rare opportunity to quantitatively study the physics and kinematics of RPS acting on an exceptionally massive late-type galaxy at the relatively high redshift of $z\sim0.3$. Viewed edge-on, and exhibiting a debris trail that extends over 40 kpc in projection on the plane of the sky, A1758N\_JFG1 would be an even more spectacular sight from a different angle, as evidenced by the system's radial peculiar velocity of close to --1000 km s$^{-1}$ and the unambiguous kinematic signature of stripped ISM trailing behind the galaxy as it approaches along our line of sight (Fig.~\ref{fig:vrotmap}). The fortuitous orientation and direction of motion of A1758N\_JFG1 provide an edge-on view of the stripping process that minimizes projection effects and clearly reveals the dynamics of both galaxy and stripped ISM. North-South asymmetries clearly visible in both optical imaging (Fig.~\ref{fig:a1758n-jfg1}) and the distribution of [OII]3727\AA\ flux (Fig.~\ref{fig:o2flux}) strongly suggest that our target's velocity vector, as seen in projection on the plane of the sky, points W or, more likely, WSW.

\subsection{ISM kinematics}

The spatial distribution of line emission from A1758N \_JFG1, specifically [\ion{O}{2}]3727\AA, provides crucial \mbox{insights} into the physics at work in this extreme case of ram-pressure stripping.  The continuous and monotonic change of both [\ion{O}{2}] flux and velocity gradient, shown in Figs.~\ref{fig:o2flux} and \ref{fig:vrotmap}, rules out the (remote) possibility of the enhanced star-formation rate and morphology of this galaxy being caused by a merger instead of RPS. More importantly, it shows that, consistent with the findings of studies of RPS in nearby galaxies \citep[e.g.,][]{bellhouse17}, stripping proceeds gradually from the outer edge of the disk inward. Although results from a single object obtained at only modest spatial resolution are unable to settle the question, the observed distribution of [\ion{O}{2}] emission also strongly suggests that the stripped ISM does not start forming stars only within the ``tentacles". In other words, enhanced star formation is triggered as ram pressure compresses the ISM in the galaxy-ICM interface, but continues far into the galaxy's wake, as gas clouds stripped from the disk collapse and form stars \textit{in situ}. Although at this early stage of the stripping process much of the [\ion{O}{2}] flux, and hence of the star formation, is still concentrated within, or just downstream of, the galactic disk \citep[in agreement with][]{kronberger08}, the balance is likely to shift more heavily toward the wake, as stripping of the ISM proceeds from the outer edges of the disk toward the galaxy's bulge.

The ISM within the disk exhibits a rotation velocity profile consistent with the TFR prediction of a maximal velocity of approximately 300 km s$^{-1}$. However, the significant line-of-sight velocity of A1758N\_JFG1 adds a characteristic tilt to a velocity gradient that would otherwise run perpendicular to the plane of the galactic disk (Fig.~\ref{fig:vrotmap}). This slant, more pronounced toward the edges of the disk, also increases with downstream distance, indicative of drag exerted by the ICM that causes the stripped gas to gradually ``fall behind'' the galaxy as both move toward us. 

\subsection{Nuclear activity}

Some of the H$\alpha$ luminosity recorded in our long-slit observation of A1758N\_JFG1 (which probed the galaxy's nucleus) could conceivably be due to nuclear activity, rather than star formation; if so, our target would be expected to harbor a prominent X-ray point source, but no such emission is detected in the existing Chandra data. The almost complete absence of signs of nuclear activity in A1758N\_JFG1 (except for the system's location in the ``composite region" of the BPT diagram presented by EK19) is intriguing, given that well established scaling relations \citep[e.g.,][]{reines15} predict the presence of a supermassive black hole (M$_{\rm BH}>10^8$ M$_\odot$) in this extreme spiral galaxy, and in view of the results of \citet{pog17} who found RPS to boost nuclear activity in nearby galaxies of high stellar mass ($M_\ast \ge 4\times 10^{10}$ M$_\odot$). 
 We note, however, that it remains entirely possible that a heavily obscured AGN resides within the very core of A1758N\_JFG1, a region that is still largely unaffected by RPS and which could be penetrated only by hard X-rays \citep[$>10$ keV; see][for a review]{hickox18}.

\subsection{Origin and trajectory}
EK19 propose that the RPS candidates identified within A1758N share their large-scale origin and trajectories with the merging subclusters, i.e., that they enter the cluster environment along a filament that is roughly aligned with the merger axis. In this scenario, the high negative radial velocities of EK19's RPS candidates require the merger axis to be substantially inclined with respect to the plane of the sky. We qualitatively tested this hypothesis with the help of a simple three-body model of the two merging subclusters and A1758N\_JFG1. While far from exhaustive, our simulations are not only able to reproduce the observational data; they do so for an orientation of the merger axis that is fully consistent with the geometry proposed by EK19 (Fig.~\ref{fig:galpy}) and disfavor the accretion of A1758N\_JFG1 directly from the field. Although our findings thus support the notion advanced by EK19 that studies of the location and apparent direction of motion of RPS events could become powerful tools to reveal the three-dimensional geometry of cluster collisions, much more sophisticated simulations (far beyond the scope of this paper) are required to fully test the potential of ``jellyfish" galaxies as diagnostics of cluster mergers.

\section{Summary}
\label{sec:summary}
A1758N\_JFG1, a newly discovered textbook ``jellyfish" galaxy, offers a rare, clean view of ram-pressure stripping at the relatively high redshift of $z=0.28$, thanks to the system's edge-on orientation. Adding to the findings of EK19 who report a dramatically enhanced star formation rate, our IFU study of the [\ion{O}{2}]3727\AA\ emission line provides a spatially resolved view of the kinematics of the ISM from the galaxy's disk to the end of the visible debris trail. We find unambiguous evidence of a sequence of gradual stripping from the outer edges of the galaxy and drag from the ICM that causes the stripped material to fall behind A1758N\_JFG1 along its path through the A1758N merger. The distribution of [\ion{O}{2}] emission strongly suggests that the enhanced star formation is the result of both ram-pressure-induced compression at the ISM/ICM interface and subsequent (and independent) \textit{in situ} collapse of ISM clouds in the debris trail.

A1758N\_JFG1 exhibits no evidence of nuclear activity, in spite of expectations of a supermassive black hole at its center based on scaling relations. Given the, at this early stage of the stripping process, potentially still very high density of gas obscuring the very core of the system, the absence of observable nuclear actively is, however, not in conflict with previous studies suggesting that RPS boosts AGN activity \citep{pog17}.

Finally, simplified modelling of the motion of A1758N\_JFG1 within the A1758N cluster merger as a three-body problem finds trajectories consistent with EK19's hypothesis of RPS candidates (including A1758N\_JFG1) being accreted by the forming cluster from a direction that is aligned with the merger axis and inclined by approximately 60$^\circ$ with respect to the plane of the sky, contrary to the commonly adopted view of the A1758N cluster merger proceeding largely in the plane of the sky \citep{machado15,mo17}. Although these results are far from definitive, given the simplicity of our simulations, they lend credence to EK19's proposal to investigate the usefulness and power of the spatial and peculiar-velocity distribution of RPS candidates as a tool to study the 3D geometry and dynamics of cluster collisions.

Having demonstrated the feasibility and power of IFU observations of extreme RPS events at intermediate redshifts, our study bodes well for a future, comprehensive kinematic analysis of the environmental dependencies of RPS in truly massive galaxy clusters. By extending the redshift range of in-depth RPS studies, for instance by targeting RPS events in MACS clusters \citep{ebeling01,ebeling14}, we will also increasingly probe dynamically young and actively evolving systems and will eventually reach the era when truly massive clusters only began to form, currently explored only by the recent study of \citet{boselli19} at $z=0.7$.

\acknowledgments
BK thanks the Sheila Watumull Astronomy Fund for financial support as a visiting researcher at IfA. We would also like to thank Conor McPartland, Emanuele Daddi, and Ivan Delvecchio for their valuable advice, as well as an anonymous referee whose insightful comments significantly improved this paper. Finally, we extend our gratitude to James Lane for his assistance with running \textsc{GALPY}.

\end{document}